\documentclass[proceedings, preprint]{rmaa}

\SetYear{2011}
\SetConfTitle{Fourth Science with the GTC}

\title{Spectroscopy of lensing galaxies in the GTC era} 

\author{
  Vyacheslav N. Shalyapin\altaffilmark{1} 
  and Luis J. Goicoechea\altaffilmark{2}}

\altaffiltext{1}{Institute for Radiophysics and Electronics, 
National Academy of Sciences of Ukraine, 12 Proskura St., 61085 Kharkov, 
Ukraine (vshal@ukr.net).}

\altaffiltext{2}{Departamento de F\'\i sica Moderna, Universidad de 
Cantabria, Avda. de Los Castros s/n, 39005 Santander, Spain 
(goicol@unican.es).}

\shortauthor{Shalyapin \& Goicoechea}
\shorttitle{Spectroscopy of lensing galaxies in the GTC era}

\listofauthors{V. N. Shalyapin \& L. J. Goicoechea}
\indexauthor{Shalyapin, V. N.}
\indexauthor{Goicoechea, L. J.}

\abstract{We are using OSIRIS to obtain spectra of faint galaxies close 
to multiply imaged quasars. We initially focused on the fields of HE 
1413+117 (Cloverleaf quasar) and SDSS 1116+4118. In this contribution, 
we present long-slit spectroscopy of two galaxies in the southwest of 
the Cloverleaf, and show that one of them makes a negligible 
contribution to the external shear of the gravitational lens system. 
Spectra of the main lensing galaxy candidate in SDSS 1116+4118 are also 
analysed and discused. If gravitational lensing is causing the quasar 
image splitting, our spectra reveal that the main lens can not consist 
of only one dominant galaxy.}

\resumen{Estamos usando OSIRIS para obtener espectros de galaxias 
d\'ebiles y pr\'oximas a cu\'asares m\'ultiples, e inicialmente nos 
hemos concentrado en los campos de HE 1413+117 (Hoja de Tr\'ebol) y SDSS 
1116+4118. En esta contribuci\'on, presentamos observaciones de dos 
galaxias cercanas a la Hoja de Tr\'ebol, mostrando que una de ellas no 
contribuye a la cizalla externa del sistema lente gravitatoria. 
Tambi\'en analizamos y discutimos espectros de la galaxia candidata a 
lente principal en SDSS 1116+4118. Si SDSS 1116+4118 es un sistema 
lente, nuestros espectros indican que la lente principal no puede estar 
formada por una \'unica galaxia dominante.}

\addkeyword{Galaxies: general}
\addkeyword{Gravitational lensing: strong}
\addkeyword{Techniques: spectroscopic}

\begin{document}
% Typeset article header
\maketitle

\section{First targets and goals}
\label{sec:intro}

Time delays between multiple images of lensed quasars can be used in 
lensing mass reconstructions and the determination of cosmological 
parameters (e.g., \citealp{2007ApJ...660....1O}). However, a 
spectroscopic follow-up of lensing galaxies is essential to use time 
delays to constraint lensing mass distributions or cosmology (for a 
recent review see \citealp{2006glsw.conf.....M}). Here, we present the 
first observations of our spectroscopic programme with OSIRIS on the 
GTC, which are part of an ambitious project with the new 10.4 m 
telescope \citep{2003RMxAC..16..317U}. These 2011 data of faint galaxies 
at relatively large angular separations from lensed quasars have been
obtained in average seeing conditions or even worse. 

We have recently measured the three time delays of the quadruple quasar 
HE 1413+117 (Cloverleaf quasar; \citealp{2010ApJ...708..995G}). The 
external shear strength in this lens system is $\gamma \sim$ 0.1, and 
the shear direction points towards the three field galaxies 1, 2, and H2
\citep{2009ApJ...699.1578M, 2010ApJ...708..995G}. Thus, we have obtained
OSIRIS/GTC spectra of the galaxies 1 and 2, which allow us to discuss 
the contribution of these targets to the external shear on the 
Cloverleaf. We have also taken long-slit spectra of the galaxy between 
the pair of quasars SDSS 1116+4118 at redshift $z \sim$ 3. Both quasars 
are separated by 13\farcs{}7, so we want to check if the galaxy produces 
an image splitting $\Delta \theta \sim$ 14\arcsec\ for a source at $z 
\sim$ 3 \citep{2007astro.ph..1537S, 2007MNRAS.378..801E}.  

\section{Results and discussion}
\label{sec:resdis}

\subsection{HE 1413+117 (Cloverleaf quasar)}
\label{sec:clover}

\begin{figure}[!t]
  \vspace{0pt}
  \includegraphics[width=\columnwidth]{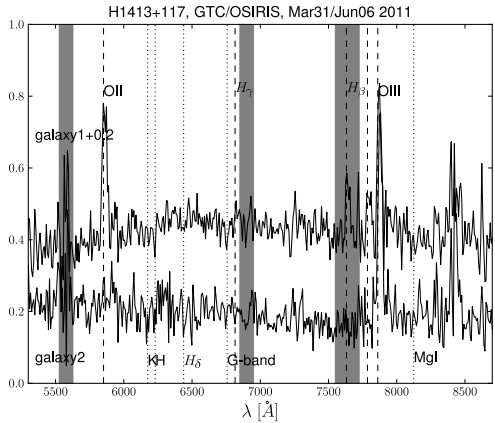}
  \caption{Spectra of the galaxies 1 and 2 around the Cloverleaf quasar. 
  Vertical dashed and dotted lines correspond to emission and absorption 
  lines at $z$ = 0.57, respectively. The three gray highlighted regions 
  are associated with prominent atmospheric and night sky features.}
  \label{fig:clover}
\end{figure}

This year we have performed seven 2 ksec spectroscopic observations of 
the galaxies 1 and 2 in the southwest of the Cloverleaf quasar, using 
the low-resolution grism R300R ($\lambda/\Delta \lambda$ = 348) and 
placing the slit along the line joining both galaxies (e.g., 
\citealp{2009ApJ...699.1578M}). From the first five datasets of each 
galaxy corresponding to March 31 and June 6 (mean seeing of $\sim$ 
1\farcs{}5), we infer the combined spectra that appear in 
Figure~\ref{fig:clover}. The galaxy No. 2 ($R \sim$ 24 mag) has a noisy
spectrum that only includes one strong emission line in the 5000--8000 
\AA\ region. Fortunately, the spectrum of the galaxy No. 1 ($R \sim$ 23 
mag) shows two strong emission lines that we identify with the 
\ion{O}{ii}($\lambda$3727) and \ion{O}{iii}($\lambda$4959, 5007), 
leading to $z$ = 0.57 $\pm$ 0.01.

Emission line galaxies at $z \sim$ 0.5--0.6 have an average rotation 
velocity of 175 km s$^{-1}$. This $V_{\mathrm{rot}}$ is also measured in 
some relatively bright ($R \sim$ 21 mag) galaxies at $z \sim$ 0.5--0.6 
(e.g., \citealp{2005MNRAS.361..109B}). Therefore, we consider a singular
isothermal sphere (SIS) model and an upper limit $V_{\mathrm{rot}} \leq$ 
175 km s$^{-1}$ to describe the halo mass of the galaxy No. 1. If the 
main lens redshift was $z$ = 0.57, then the shear from this galaxy would
be less than 0.02. However, the main lens redshift is likely about 2 
\citep{2010ApJ...708..995G}, and thus, the effective shear should be 
less than 0.002 \citep{2006ApJ...641..169M}. This shear strength is 
significantly lower than the external shear in the system ($\gamma \sim$ 
0.1).

\subsection{SDSS 1116+4118}
\label{sec:candi}

\begin{figure}[!t]
  \vspace{0pt}
  \includegraphics[width=\columnwidth]{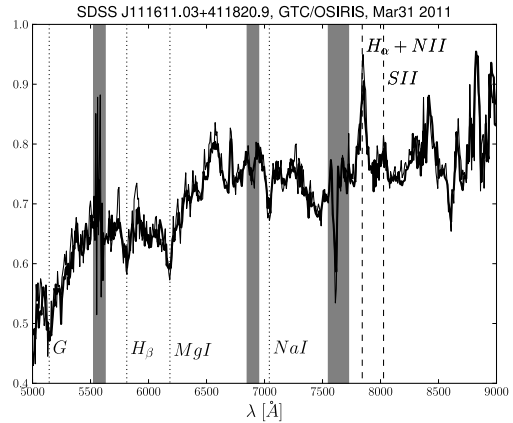}
  \caption{Spectra of the candidate for main lensing galaxy of SDSS 
  1116+4118. The darkest line represents the spectrum along the major 
  axis of the galaxy, while the other line describes the spectrum along 
  its minor axis. The gray highlighted regions have the same meaning as 
  in Fig. 1, and the vertical marks correspond to emission and 
  absorption lines at $z$ = 0.195.}
  \label{fig:candi}
\end{figure}

We have observed the lens system candidate on March 31 and April 7, 
2011. The low-resolution spectra along the major and minor axes of the 
galaxy between the pair of quasars at $z \sim$ 3 are displayed in 
Figure~\ref{fig:candi}. The exposure times/seeing values are 2.0 
ksec/2\farcs{}00 (major axis) and 1.2 ksec/2\farcs{}35 (minor axis), and 
both spectra are consistent with each other. From the Kinney-Calzetti 
templates, we conclude that the candidate for main lensing galaxy ($R 
\sim$ 19 mag; \citealp{2007astro.ph..1537S}) is a spiral (Sb?) at $z$ = 
0.195 $\pm$ 0.002. 

Assuming a SIS model and a standard rotation velocity ($V_{\mathrm{rot}} 
\leq$ 300 km s$^{-1}$), the lensing galaxy candidate can only generate 
an image splitting $\Delta \theta \leq$ 2\arcsec\ for a source at $z 
\sim$ 3. This is well below the observed separation between the pair of 
quasars. Even for an extremely massive spiral with $V_{\mathrm{rot}} 
\sim$ 500 km s$^{-1}$, e.g., UGC 12591 \citep{1986ApJ...301L...7G} or 
ISOHDFS 27 \citep{2002ApJ...580..789R}, the predicted image splitting is
$\Delta \theta \sim$ 6\arcsec. This extreme value is less than half of
the angular separation between the two distant quasars. 

Based on observations made with the Gran Telescopio Canarias (GTC), 
instaled in the Spanish Observatorio del Roque de los Muchachos of the 
Instituto de Astrofísica de Canarias, in the island of La Palma.

\end{document}